\documentclass[10pt]{iopart}
\usepackage{graphicx}
\usepackage{subfigure}
\usepackage{verbatim}
\usepackage{dcolumn}
\usepackage{bm}
\usepackage{epsf}
\usepackage{color}
\usepackage[toc]{appendix}
\usepackage{stmaryrd}
\usepackage{amsthm}
\usepackage{hhline}
\usepackage{amssymb}
\usepackage{iopams}
\usepackage{cite}
\usepackage{tikz}
\usetikzlibrary{arrows,shapes,calc,matrix}

\expandafter\let\csname equation*\endcsname\relax
\expandafter\let\csname endequation*\endcsname\relax
\usepackage{amsmath}
\usepackage{xstring}

\makeatletter
\newcommand\footnoteref[1]{\protected@xdef\@thefnmark{\ref{#1}}\@footnotemark}
\makeatother

\newcommand\invisiblesection[1]{%
	\refstepcounter{section}%
	\addcontentsline{toc}{section}{\protect\numberline{\thesection}#1}%
	\sectionmark{#1}}

\newcommand{\la}{\left\langle}
\newcommand{\ra}{\right\rangle}
\newcommand{\pd}{\partial}

\newcommand{\ex}[1]{\exp{\left(#1\right)}}
\newcommand{\loge}[1]{\ln{\left(#1\right)}}

\newcommand{\bla}{bla\\bla\\bla\\bla\\bla}
\usepackage{fmtcount}

\newcommand{\mb}[1]{\mbox{\boldmath$#1$}}

\newcommand{\mrm}[1]{\mathrm{#1}}

\renewcommand{\appendix}{
}

\newcommand{\draftmode}{1}    
\newcommand{\notetoself}[1]{\ifnum \draftmode=1 {\color[rgb]{0,0,0.8} [#1]} \fi}  
\newcommand{\cuttext}[1]{\ifnum \draftmode=1 {\color[rgb]{0,0.5,0} [#1]} \fi}  
\newcommand{\warntext}[1]{\ifnum \draftmode=1 {\color[rgb]{0.9,0.6,0} #1} \else {#1} \color{black} \fi}
\newcommand{\aref}[1]{{Appendix~\hyperref[#1]{A}}}
\newcommand{\bref}[1]{{Appendix~\hyperref[#1]{B}}}
\usepackage[colorlinks=true,citecolor=blue,linkcolor=blue,urlcolor=blue]{hyperref}
\usepackage{doi}
\usepackage{cleveref}

\begin{document}

\title{Stochastic thermodynamics of relativistic Brownian motion}

\author{P. S. Pal\textsuperscript{1} and Sebastian Deffner\textsuperscript{2}}
\ead{$^1$pspal@umbc.edu}
\ead{$^2$deffner@umbc.edu}
\address{Department of Physics, University of Maryland, Baltimore County, Baltimore, MD 21250, USA}

\begin{abstract}
Physical scenarios that require a relativistic treatment are ubiquitous in nature, ranging from cosmological objects to charge carriers in Dirac materials. Interestingly all of these situations have in common that the systems typically operate very far from thermal equilibrium. Therefore, if and how the framework of stochastic thermodynamics applies at relativistic energies is a salient question. In the present work we generalize the notions of stochastic heat and work for the relativistic Langevin equation and derive the fluctuation theorems without and with feedback. For processes with feedback we consider the ramifications of the lack of simultaneity of events in the inertial frames of observer and Brownian particles, and we argue that the framework of absolute irreversibility is instrumental to avoid acausal considerations. The analysis is concluded with a few remarks on potential experimental applications in graphene.
\end{abstract}

\section{Introduction}

The Universe is full of objects that defy the common anthropocentric understanding of nature. A good example are so-called blazars, which are caused by narrow jets of cosmic particles moving at relativistic speeds directly towards Earth \cite{Meyer2018}. To date, only very little is known about these jets since ``most of the action'' occurs over length scales that are too small to be resolved even with the best telescopes. Curiously, what is known \cite{Meyer2018}, is that these jets are powered by supermassive black hole ``engines'' that propel charged particles into the Universe.

From a thermodynamic point of view, these cosmic jets are, thus, nothing else but ensembles of Brownian particles that are driven very far from thermal equilibrium. Therefore, it seems plausible that recent advances in stochastic thermodynamics \cite{Sekimoto1998,Seifert2005,Seifert2008,Seifert2012,Broeck2016,Yang2020} should be useful in describing the thermal properties also of cosmic jets. However, with the exception of the fluctuation theorem derived by Fingerle \cite{Fingerle2007}, none of the existing results appear directly applicable at relativistic energies.

Fluctuation theorems are statements of the second law of thermodynamics that quantify the occurrence of negative fluctuations of the entropy \cite{Evans1993,Gallavotti1995,Jarzynski1997,Crooks1999}. Over the last three decades these fluctuation theorems have proven themselves as robust findings that hold for a wide range of systems driven arbitrarily far from equilibrium \cite{Evans2002}, that have also been generalized to the quantum domain \cite{Campisi2009,Campisi2011,Talkner2016,Bartolotta2018,Deffner2019book}, and which have been experimentally tested \cite{Liphardt2002,Collin2005,Batalhao2014,An2015,Ciliberto2017,Smith2018,Gardas2018}. Moreover, they inspired the rapid development of ``thermodynamics of information'' \cite{Sagawa2008,Sagawa2010,Barato2014,Parrondo2015,Strasberg2017,Esposito2019,Wolpert2019}, since analyzing thermodynamic properties along trajectories of single particles is reminiscent of the original mindset of Maxwell's demon \cite{Mandal2012,Mandal2013,Deffner2013,Deffner2013PRE,Boyd2016,Safranek2018,Stevens2019}.

The purpose of the present paper is to generalize the framework of stochastic thermodynamics to relativistic Brownian motion. For the sake of simplicity, we will focus on the relativistic Ornstein-Uhlenbeck process \cite{Debbasch1997,Debbasch1998,Barbachoux2001,Barbachoux2001_2}, for which the Langevin equation has been extensively studied \cite{Dunkel2005,Dunkel2005PRE2,Dunkel2009,Dunkel2009PR}. As main results, we will identify the relativistic expression of the stochastic heat, work, and entropy production. For the entropy production we will further prove the detailed as well as the integral fluctuation theorems. In a second part of the analysis, we will study relativistic Brownian motion under feedback. For such a relativistic version of Maxwell's demon, we will derive a statement of the second law that is reminiscent of formulations for \emph{absolutely irreversible} processes \cite{Murashita2014}. To this end, we will elucidate how monitored Brownian motion becomes absolutely irreversible due to the relativistic time delay. The analysis will be concluded with a few remarks on potential experimental implementations in novel materials with effectively relativistic dispersion relations.

\section{\label{roup} Stochastic energetics of the relativistic Ornstein-Uhlenbeck process}

Consider a Brownian particle in one dimension with rest mass $M$ and in contact with a heat reservoir at temperature $T$. As usual, the heat bath is taken to be at rest with respect to the inertial frame of the laboratory \cite{Debbasch1997,Debbasch1998}. Then, the dynamics of the Brownian particle is described by the relativistic Langevin equation 
\begin{equation}
\label{eq:rel_Lang}
\dot x =\frac{p}{p^0}\quad\text{and}\quad\dot p=-\gamma M\,\frac{p}{p^0}-\pd_x\,U(x,\alpha(t))+\sqrt{2D}\,\xi(t)\,,
\end{equation}
where $x$ is the position, $p$ denotes the canonical momentum, and $p^0=\sqrt{p^2+M^2}$ is the relativistic kinetic energy. Note that to avoid clutter in the formulas we work in units such the speed of light $c=1$. Further, $\gamma$ is the viscous friction coefficient, and $D$ is the diffusion coefficient for the Gaussian white noise $\xi(t)$, with $\la\xi(t)\ra=0$ and $\la\xi(t)\xi(t')\ra=\delta(t-t')$. The noise is taken to be Gaussian and white merely for the sake of simplicity. For more involved scenarios we refer to the literature \cite{Dunkel2009PR}. Finally, $U(x,\alpha(t))$ is the time-dependent potential and $\alpha(t)$ is an external control parameter.

 It is interesting to note that generally Eq.~\eqref{eq:rel_Lang} is not covariant. This rest in the fact that any theory based on observer-simultaneously defined quantities necessarily leads to a loss of covariance \cite{Dunkel2009PR}. In principle, the issue could be avoided by formulating a thermodynamic field theory. For the sake of simplicity and accessibility, however, we have chosen to work with the relativistic Langevin equation \eqref{eq:rel_Lang}, which has been thoroughly developed by Dunkel and H\"anggi \cite{Dunkel2005,Dunkel2005PRE2,Dunkel2009,Dunkel2009PR}.

 In this framework it is then a simple exercise to generalize stochastic energetics \cite{Sekimoto1998} to the relativistic Langevin equation \eqref{eq:rel_Lang}. To this end, consider the relativistic energy-momentum relation,
\begin{equation}
\label{eq:E}
E(x,\alpha(t))=\sqrt{p^2+M^2}+U(x,\alpha(t))\,,
\end{equation}
from which we obtain 
\begin{equation}
\label{eq:1stlaw}
\frac{dE}{dt}=\frac{p}{p^0}\left(-\gamma M\,\frac{p}{p^0}+\sqrt{2D}\,\xi(t)\right)+\frac{\partial U}{\partial \alpha}\,\dot\alpha\,.
\end{equation}
Inspecting Eq.~\eqref{eq:1stlaw} we immediately recognize the thermodynamic work for a process of length $\tau$ \cite{Sekimoto1998,Horowitz2007}
\begin{equation}
\label{eq:W}
W=\int_{0}^{\tau}dt\,\frac{\partial U}{\partial \alpha}\,\dot\alpha\,,
\end{equation}
which is the identical expression as found in non-relativistic Brownian motion. Accordingly, the heat becomes
\begin{equation}
\label{eq:Q}
Q=\int_{0}^{\tau}dt\,\frac{p}{p^0}\left(-\gamma M\,\frac{p}{p^0}+\sqrt{2D}\,\xi(t)\right)\,.
\end{equation}
Note that both heat, $Q$, and work, $W$, explicitly depend on the fluctuating position and momentum, and thus also $Q$ and $W$ are stochastic quantities. Here and in the following, we tacitly assume the standard, Stratonovich interpretation for the stochastic integrals \cite{Bo2019,Lim2019}.

It is interesting to note that in the non-relativistic limit, i.e, if the speed of the Brownian particle is much smaller than the speed of light,  $p^0$ simply reduces to $p^0=M$. Therefore, we recover the classical expression for the stochastic heat in the non-relativistic limit.

The relativistic expressions for work and heat in Eqs.~\eqref{eq:W} and \eqref{eq:Q}, respectively, constitute our first main result. It is interesting to note that the expression for the stochastic work is identical to the non-relativistic case. Work is commonly understood as an ``ordered'' exchange of internal energy \cite{Callen1985}. Thus, from the point of view of the Brownian particle, how to compute work should not depend on how fast the particle is traveling. Heat, on the other hand, is the energy ``lost'' into the heat bath, which in the present scenario is at rest with respect to the Brownian particle. Microscopically, heat is exchanged through random interaction of the Brownian particle with the degrees of freedom of the bath. Thus, the relativistic description of momentum transfer is pertinent, and one would already intuitively expect that the expression for heat does depend on the relative motion.

\section{Stochastic thermodynamics and fluctuation theorems}

We now continue the analysis by focusing on the relativistic entropy production. To this end, we will first identify the internal and external contributions by comparing with the stochastic heat \eqref{eq:Q}, before we will proof the fluctuation theorems.

\subsection{Relativistic entropy production}

The stochastic entropy production is more conveniently described through the time-evolution of the phase space density $\rho(\Gamma,t)$ \cite{Jarzynski1997PRE,Seifert2005,Esposito2010,Esposito2010PRE}. Thus, we consider the  relativistic Fokker Planck equation \cite{Dunkel2005}
\begin{equation}
\label{eq:FP}
\partial_t\,\rho(\Gamma,t)=-\nabla\cdot\mb{j}(\Gamma,t)
\end{equation}
where $\Gamma=(x,p)$ is a point in phase space, $\nabla\equiv(\partial_x,\partial_p)$, and $\mb{j}(\Gamma,t)$ is the probability current. Under time-reversal, the position, $x$, is an even function whereas the momentum, $p$, is an odd function. Thus, it has proven convenient to separate $\mb{j}$ into two contributions, $\mb{j}\equiv\mb{j}_\mrm{r}+\mb{j}_\mrm{d}$ \cite{Esposito2010PRE}. We have
\begin{equation}
\mb{j}_{\text{r}}=\left(\frac{p}{p^0},\,-\partial_x U(x,\alpha(t))\right)\,\rho(\Gamma,t)\,,
\end{equation}
which is symmetric under time-reversal, and a dissipative part
\begin{equation}
\label{eq:jd}
\mb{j}_{\text{d}}=\left(0,-\gamma M\,\frac{p}{p^0}-D\,\partial_p\right)\,\rho(\Gamma,t)\,,
\end{equation}
that is asymmetric under time-reversal. It is this second contribution that is the origin of irreversible entropy production \cite{Esposito2010,Esposito2010PRE}.

\subsubsection{Relativistic fluctuation-dissipation theorem and detailed balance}

Before analyzing the entropy production we need to consider the proper definition of thermodynamic temperature. To this end, consider that in equilibrium for a specific value of $\alpha$ the dissipative part of the current \eqref{eq:jd} vanishes, $\mb{j}_{\text{d}}=0$, and therefore we also have $\nabla \cdot\mb{j}_{\text{r}}=0$. 

Accordingly, the stationary solution of \eqref{eq:FP} can be written as,
\begin{equation}
\label{eq:rhoeq}
\rho^\mrm{eq}(\Gamma,\alpha)=\rho_0\,\ex{-\frac{\gamma M}{D}\,E(x,\alpha)}\,,
\end{equation}
where $E(x,\alpha)$ is again the relativistic energy \eqref{eq:E}. Now demanding that the equilibrium state \eqref{eq:rhoeq} is given by the Maxwell-J\"uttner distribution \cite{Juttner1911,Dunkel2005} we obtain the corresponding fluctuation-dissipation theorem
\begin{equation}
\label{eq:flucdiss}
D=\frac{\gamma M}{k_B T}\,,
\end{equation}
which is identical to the non-relativistic case.

\subsubsection{Second law of relativistic stochastic thermodynamics}

Now, consider the non-equilibrium Shannon entropy of the Brownian particle
\begin{equation}
S(t)=-k_B\int d\Gamma\, \rho(\Gamma,t)\loge{\rho(\Gamma,t)} \,,
\end{equation}
whose rate can be written as
\begin{equation}
\dot S(t)=-k_B\int d\Gamma\, \frac{\partial \rho}{\partial t} \loge{\rho(\Gamma,t)}=k_B\int d\Gamma\, \nabla\cdot\left(\mb{j}_{\text{r}}+\mb{j}_{\text{d}}\right)\,\loge{\rho(\Gamma,t)}\,\,.
\end{equation}
Using standard arguments \cite{Esposito2010PRE} it is easy to see that the contribution from the time-reversal current, $\mb{j}_{\text{r}}$ vanishes. 

Thus, we have after integrating by parts twice and simplifying expressions
\begin{equation}
\label{eq:sdot}
\dot S(t)=k_B\int d\Gamma \frac{\rho}{D}\left(\frac{\mb{j}_\mrm{d}}{\rho}\right)^2+k_B\,\frac{\gamma M}{D}\int d\Gamma\, \frac{p}{p^0}\,j_\mrm{d}^p\,.
\end{equation}
and where we introduced the notation $j_\mrm{d}^p=-\gamma M  p/p^0\,\rho-D\,\partial_p\rho$.

To continue we now need to identify the contribution to the entropy production that is solely due to the exchange of heat between the Brownian particle and the thermal reservoir. Thus, inspecting the second term in Eq.~\eqref{eq:sdot} we define,
\begin{equation}
\dot s_\mrm{res}\equiv-k_B\frac{\gamma M}{D}\, \frac{p}{p^0}\,\frac{j_\mrm{d}^p}{\rho} \quad\text{with}\quad \dot{S}_\mrm{res}=\int d\Gamma\, \dot s_\mrm{res}\,\rho(\Gamma,t)\,.
\end{equation}
Now re-writing the momentum component of the total probability current, $\mb{j}(\Gamma,t)=(j^x,j^p)$, as
\begin{equation}
j^p=\left(-\frac{\gamma M p}{p^0}-\partial_x U-D\partial_p\right)\rho(\Gamma,t)=-\partial_x U\,\rho(\Gamma,t)+j_\mrm{d}^p
\end{equation}
we have
\begin{equation}
\Delta s_{\text{res}}=\int_0^\tau dt\,\dot s_\mrm{res}=-k_B\,\frac{\gamma M}{D}\int_0^{\tau}dt\,\left(\dot p+\partial_x U\right)\frac{p}{p^0}\,,
\end{equation}
where we also used that $\dot p=j^p/\rho$.

Finally, employing the relativistic Langevin equation \eqref{eq:rel_Lang} we can write,
\begin{equation}
\label{eq:entres}
\Delta s_{\text{res}}=-\frac{1}{T}\int_0^{\tau}dt\,\left(-\gamma M\,\frac{p}{p^0}+\sqrt{2D}\,\xi(t)\right)\frac{p}{p^0}=-\frac{Q}{T} \,,
\end{equation}
where we identified the stochastic heat \eqref{eq:Q} substituted the fluctuation-dissipation theorem \eqref{eq:flucdiss}. Therefore, the total irreversible entropy production becomes
\begin{equation}
\label{eq:entprod}
\dot S_{\text{tot}}(t)=\dot S(t)+\dot S_{\text{res}}(t)=\frac{k_B}{D}\,\int d\Gamma\,\frac{\mb{j}^2_d}{\rho}\ge 0\,,
\end{equation}
which is always non-negative and of the standard form \cite{Esposito2010PRE}.

Equation~\eqref{eq:entprod} is a statement of the second law of thermodynamics for relativistic Brownian motion. It states that the irreversible entropy production is always non-negative, and thus Eq.~\eqref{eq:entprod} can also be considered a relativistic generalization of the Clausius inequality.

\subsection{Fluctuation theorem for the irreversible entropy production}

Finally, we need to show that the fluctuation theorem also remains valid at relativistic energies. To this end, we will generalize the path integral formalism pioneered in Refs.~\cite{Chernyak2006,Deffner2011EPL}. As before we consider a relativistic Brownian particle that is driven by the time-dependent external control parameter $\alpha(t)$. Moreover, we denote a trajectory in phase space by $\mb{X}_t\equiv\{\Gamma(t)\}$.

Now, consider a sequence of time steps of length $\delta t$ with $N\delta t=\tau$, such that the trajectory $\mb{X}_t$ is separated into $N$ infinitesimal segments. Then,  $\rho^+_i(\Gamma_{i+1},t_i+\delta t|\Gamma_i,t_i)$ is the transition probability between two consecutive points along  $\mb{X}_t$. It can be written as
\begin{equation}
\rho_i^+=J_{\xi_i,p_i}^+\,\la\delta\left(\dot x_i-\frac{p_i}{p^0}\right)\delta\left(\dot p_i+\frac{\gamma M}{p^0} p_i+\partial_{x_i}U_i-\sqrt{2D}\xi_i\right)\ra_{\xi_i},
\end{equation}
where $\la...\ra_{\xi_i}$ is an average over the noise with probability distribution
\begin{equation}
P(\xi_i)=\sqrt{\frac{\delta t}{2}}\,\ex{-\frac{\delta t}{2}\xi_i^2}\,,
\label{noise}
\end{equation}
and $J_{\xi_i,p_i}^+$ is the Jacobian. In the present case, the Jacobian becomes,
\begin{equation}
J_{\xi_i,p_i}^+=\det(\partial_{p_i}\xi_i)=\frac{1}{\delta t\sqrt{2D}}\left[1-\frac{\delta t}{2}\gamma M\,\partial_{p_i}\left(\frac{p_i}{p^0}\right)\right].
\label{j+}
\end{equation}
which follows from straight forwardly evaluating terms.

Using the explicit expression for the noise distribution \eqref{noise} the transition probability for the forward process reads
\begin{equation}
\label{eq:transprob}
\rho_i^+=J_{\xi_i,p_i}^+\,\sqrt{\frac{\delta t}{4\pi D}}\,\delta\left(\dot x_i-\frac{p_i}{p^0}\right)\ex{-\frac{\delta t}{4D}\left(\dot p_i+\frac{\gamma M}{p^0} p_i+\partial_{x_i}U_i\right)^2}\,.
\end{equation}
The corresponding transition probability for the reverse process, i.e., the transition probability for the time-reversed driving becomes 
\begin{equation}
\rho_i^-=J_{\xi_i,p_i}^-\sqrt{\frac{\delta t}{4\pi D}}\,\delta\left(\dot x_i-\frac{p_i}{p^0}\right)\ex{-\frac{\delta t}{4D}\left(\dot p_i-\frac{\gamma M}{p^0} p_i+\partial_{x_i}U_i\right)^2},
\label{f-}
\end{equation}
where $J_{\xi_i,p_i}^-$ is the reverse Jacobian. Note that the reverse Jacobian is identical to the Jacobian of the forward process, since it only contains even functions of the momentum \cite{Onsager1953}.

We are now equipped with everything we need to compute the ratio of the transition probabilities for forward and reverse process. We have,
\begin{equation}
\label{eq:ratio}
\begin{split}
\frac{\rho^+}{\rho^-}&=\prod_{i=1}^N\frac{\rho_i^+}{\rho_i^-}=\prod_{i=1}^N\ex{-\frac{\gamma M}{D}\,(\dot p_i+\partial_{x_i}U_i)\,\frac{p_i}{p^0}\,\delta t}\\
&=\ex{-\frac{\gamma M}{D}\int_0^{\tau}dt\, (\dot p+\partial_x U)\,\frac{p}{p^0} }=\ex{\Delta s_\text{res}/k_B}\,,
\end{split}
\end{equation}
where we used the definition of the integral in the first line, and identified the $\Delta s_\text{res}$ in the second line. Note again that in the non-relativistic limit $p^0=M$, and thus  Eq.~\eqref{eq:ratio} reduces to the classical expression for slow velocities.

Further assuming that both, forward as well as reverse process start in the respective equilibrium state $\rho_{\text{in}}^s$ and $\rho_{\text{fin}}^s$, respectively, the system entropy change is $\Delta s=k_B\ln(\rho_{\text{in}}^s/\rho_{\text{fin}}^s)$. Therefore, using $\rho^{\text{F}}=\rho_{\text{in}}^s\rho^+$ and $\rho^{\text{R}}=\rho_{\text{fin}}^s\rho^-$ we finally obtain,
\begin{equation}
\label{eq:detailedFT}
\rho^\text{F}=\ex{\Delta s_{\text{tot}}/k_B}\, \rho^\mrm{R}
\end{equation}
where as before $\Delta s_{\text{tot}}=\Delta s+\Delta s_{\text{res}}$. Equation~\eqref{eq:detailedFT} is the detailed fluctuation theorem for the total entropy production in relativistic Brownian motion. Accordingly, we also have an integral fluctuation theorem $\la\ex{-\Delta s_{\text{tot}}/k_B}\ra=1$.

As a main result, we have shown that with the proper identification of the relativistic work \eqref{eq:W} and heat \eqref{eq:Q} the major results of stochastic thermodynamics hold without modification also at relativistic energies.

Finally, we note that the relativistic Crooks fluctuation theorem trivially follows from Eq.~\eqref{eq:detailedFT}. To this end, consider that the total entropy production ($\Delta s_\text{tot}$) can be divided into two parts, namely the system entropy change ($\Delta s$) and the reservoir entropy change ($\Delta s_\text{res}$). The reservoir entropy change can be written in terms of heat \eqref{eq:entres} and system entropy change can be written in terms of the internal energy change (using Eq.~\eqref{eq:rhoeq}) and the free energy change. Finally using the first law  \eqref{eq:1stlaw} one obtains the work fluctuation theorem.

\section{Maxwell's demon at relativistic energies}
\label{info}

Initiated by the seminal work of Sagawa and Ueda \cite{Sagawa2008,Sagawa2010} the study of Maxwell's demon as experienced a renaissance. In the remainder of the analysis, we will now show that the fundamental postulates of special relativity inhibit the common working principles of the demon.

Einstein's central insight was that no object can be travel faster than the speed of light \cite{Einstein1905}, and hence no information can be transferred faster than light. This observation is at variance with the usual set-up of Maxwell's gedankenexperiment, in which a neat fingered being observes the physical state of gas molecules and takes \emph{immediate} action \cite{Leff2014}. If the observed particles are moving at relativistic speed relative to the observer, any feedback will be subject to the relativistic time delay. This is nothing else but a direct consequence of lack of simultaneity of events in the inertial frames of the observer and the Brownian particle. Consequently, the natural question arises how this fact manifests itself in the statements of the second law.

To gain insight, we now consider a relativistic Brownian particle, whose dynamics is described by
\begin{equation}
\label{eq:LangFB}
\dot x=\frac{p}{p^0}\quad\text{and}\quad \dot p=-\gamma M\,\frac{p}{p^0}-\partial_xU+F_{\text{fb}}[\mb{X}_{t-\tau_0}]+\sqrt{2D}\,\xi(t)\,,
\end{equation}
where $F_{\text{fb}}[\mb{X}_{t-\tau_0}]$ is the force arising from autonomous or non-autonomous feedback \cite{Aastrom2010}, which depends on the microscopic trajectory upto  time  $(t-\tau_0)$. For instance, if $F_{\text{fb}}[\mb{X}_{t-\tau_0}]$ is proportional to the position $x$ at time $(t-\tau_0)$, then $\tau_0$ is just given by standard time dilation. Interestingly, the mathematical description of relativistic Brownian motion \eqref{eq:LangFB} is analogous to physical systems with response lags \cite{Rosinberg2015}.

Accordingly, the energy balance, i.e., the first law of thermodynamics \eqref{eq:1stlaw} now reads
\begin{equation}
\label{eq:1stlawFB}
\dot E=\dot Q+\dot W+\dot W_{\text{fb}}
\end{equation}
where $\dot W_{\text{fb}}=(p/p^0)F_{\text{fb}}[\mb{X}_{t-\tau_0}]$ is the power input due to the  feedback force. Equation~\eqref{eq:1stlawFB} constitutes the generalization of relativistic stochastic energetics to systems with feedback.

More interesting are the ramifications for the second law and the fluctuation theorems. To this end, we re-consider the transition probability between two increment \eqref{eq:transprob}. We immediately conclude that
\begin{equation}
\rho_i^+=J_{\xi_i,p_i}^+\sqrt{\frac{\delta t}{4\pi D}}\,\delta\left(\dot x_i-\frac{p_i}{p^0}\right)\ex{-\frac{\delta t}{4D}\left(\dot p_i+\frac{\gamma M}{p^0} p_i+\partial_{x_i}U_i-F_{\text{fb}}[\mb{X}_{t_i-\tau_0}]\right)^2}\,,
\end{equation}
where as before the Jacobian is given by Eq.~\eqref{j+}.

The problem now is how to identify the correct reverse or adjoint process, which should be compared in the fluctuation theorem \cite{Horowitz2011}. Whereas already in the case of instantaneous feedback the question is non-trivial \cite{Horowitz2007}, the situation becomes significantly more involved at relativistic energies. Due to the relativistic time delay any direct time-reversal of the trajectories would necessarily lead to acausal events, and as such naive time-reversal cannot be considered thermodynamically sound. 

A possible solution can be found by considering the recently developed notion of absolute irreversibility \cite{Murashita2014}. Consider a system that undergoes a thermodynamic process with microscopic trajectory $\mb{X}^F$ that occurs with probability $\rho^F[\mb{X}^F]$. A time-reversed trajectory $\mb{X}^R$ may then occur with probability $\rho^R[\mb{X}^R]$. The systems exhibits absolute irreversibility if either (i) $\rho^F[\mb{X}^F]>0$ and  $\rho^R[\mb{X}^R]=0$, or  (ii) $\rho^F[\mb{X}^F]=0$ and  $\rho^R[\mb{X}^R]>0$. In other words, a process becomes absolutely irreversible if the probability to observe the reverse trajectory vanishes.

Realizing that the probability of acausal events occurring has to be zero the notion of absolute irreversibility appears well-suited to study relativistic Brownian motion. In complete analogy to the non-relativistic case \cite{Murashita2014} we therefore consider the ``reverse'' dynamics to be described by 
\begin{equation}
\label{eq:Langreverse}
\dot x=\frac{p}{p^0}\quad\text{and}\quad\dot p=\gamma M\,\frac{p}{p^0} -\partial_xU+\sqrt{2D}\,\xi(t)\,,
\end{equation} 
which is nothing else but the time-reversed dynamics of Eq.~\eqref{eq:LangFB} in the absence of any feedback.

For this scenario the ratio of forward \eqref{eq:LangFB} and reverse \eqref{eq:Langreverse} transition probabilities \eqref{eq:ratio} becomes
\begin{equation}
\frac{\rho^+}{\rho^-}=\ex{-\frac{\gamma M}{D}\int_0^{\tau} dt\, (\dot p+\partial_x U)\frac{p}{p^0}+\frac{1}{2D}\int_0^{\tau} dt \left(\dot p+\gamma M\frac{p}{p^0} \partial_x U\right)F_{\text{fb}}[\mb{X}_{t-\tau_0}]}
\end{equation}
which differs from the case without feedback \eqref{eq:ratio} only by one additional term. In complete analogy to before the probability distribution of all forward trajectories is then given by $\rho^F=\rho^+\rho^F_{\mrm{in}}$, where $\rho_{\mrm{in}}$ is again the initial distribution. Correspondingly we also have for the reverse process $\rho^R=\rho^-\rho^R_{\mrm{in}}$. Hence the detailed fluctuation theorem becomes
\begin{equation}
\rho^F=\exp(R)\,\rho^R\,,
\label{eq:dftFB}
\end{equation}
where we introduced the dissipation function $R$, which reads
\begin{equation}
\begin{split}
R&=-\frac{\gamma Mk_B}{D}\int_0^{\tau} dt\, (\dot p+\partial_x U)\frac{p}{p^0}+\ln\left(\frac{\rho_{\text{in}}^F}{\rho_{\text{in}}^R}\right)\\
&+\frac{k_B}{2D}\int_0^{\tau} dt \left(\dot p+\gamma M\frac{p}{p^0}+\partial_x U\right)F_{\text{fb}}[\mb{X}_{t-\tau_0}]\,.
\end{split}
\label{sr2}
\end{equation}
We immediately recognize the first two terms as the stochastic entropy production $\Delta s_\mrm{tot}$, and thus we can also write
\begin{equation}
\label{eq:R}
R=\Delta s_\mrm{tot}+\frac{k_B}{2D}\int_0^{\tau} dt\, \left(\dot p-\gamma M\frac{p}{p^0}+\partial_x U\right)F_{\text{fb}}[\mb{X}_{t-\tau_0}]\,,
\end{equation}
which is the total dissipation function comprised of entropy production and irreversible work input due to the feedback force, see also Ref.~\cite{Murashita2014} for the interpretation of $R$ in the context of absolutely irreversible processes.

Note that $R$ in Eq.~\eqref{eq:R} contains only those pairs of trajectories, $\mb{X}_{\tau}^F$ and $\mb{X}_{\tau}^R$, that are the time-reversal of each other. In particular, there exist some trajectories that are generated by Eq.~\eqref{eq:LangFB}, whose time-reversed version cannot be generated by the Langevin equation \eqref{eq:Langreverse}. Similarly, Eq.~\eqref{eq:Langreverse} generates trajectories that do not have a conjugate in the forward process.

Let $\Xi^F$  denote the set of all trajectories generated by the forward dynamics Eq.~\eqref{eq:LangFB} that do have a time-reversed version and $\Xi^F_c$ be the set of trajectories that do not have a time-reversed version, i.e., the complement of  $\Xi^F$. Similarly, $\Xi^R$ denotes the set of all trajectories generated by reverse dynamics Eq.~\eqref{eq:Langreverse} that have a conjugate in the forward process and $\Xi^R_c$ is the complementary set consisting of all the other trajectories that are generated by reverse dynamics. Then, we can write
\begin{equation}
\begin{split}
&\la\exp(-R)\ra=\int_{\Xi^F\cup\, \Xi^F_c}\,\mathcal{D}\mb{X}_{\tau}^F\, \exp(-R) \rho^F[\mb{X}_{\tau}^F]\\
&\quad=\int_{\Xi^R}\mathcal{D}\mb{X}_{\tau}^R\, \rho^R[\mb{X}_{\tau}^R]+\int_{\Xi^F_c}\mathcal{D}\mb{X}_{\tau}^F\, \exp(-R) \rho^F[\mb{X}_{\tau}^F]\\
&\quad=\int_{\Xi^R\cup\,\Xi^R_c}\mathcal{D}\mb{X}_{\tau}^R\, \rho^R[\mb{X}_{\tau}^R]-\int_{\Xi^R_c}\,\mathcal{D}\mb{X}_{\tau}^R \rho^R[\mb{X}_{\tau}^R]+\int_{\Xi^F_c}\,\mathcal{D}\mb{X}_{\tau}^F \exp(-R) \rho^F[\mb{X}_{\tau}^F]\,,
\label{FT1}
\end{split}
\end{equation}
where we used the detailed fluctuation theorem \eqref{eq:dftFB} in the second line.

Now, denoting the total probability of reverse trajectories that do not have any counterpart in the forward process by $\lambda\equiv\int_{\Xi^R_c}\mathcal{D}\mb{X}_{\tau}^R\, \rho^R[\mb{X}_{\tau}^R]$, and introducing  $\Lambda\equiv\int_{\Xi^F_c} \mathcal{D}\mb{X}_{\tau}^F\,\exp(-R) \rho^F[\mb{X}_{\tau}^F]$ we obtain
\begin{equation}
\label{eq:iftFB}
\la\exp(-R)\ra=1-\lambda+\Lambda\,.
\end{equation}
Equation~\eqref{eq:iftFB} constitutes a modified version of the integral fluctuation theorem due to the underlying absolutely irreversible processes, see also Ref.~\cite{Murashita2014}. Further noting that $\lambda$ and $\Lambda$ are both  positive numbers, we can write with the help of Jensen's inequality
\begin{equation}
\la R\ra\geq\ln(1-\lambda+\Lambda)\,,
\end{equation}
which is a generalized Clausius inequality for relativistic Brownian motion under feedback. This generalized Clausius inequality provides a lower bound on the dissipating function $\la R\ra$. Quite remarkably, the lower bound can become negative, which means that there exists processes for which the average dissipation is \emph{negative}. This can be understood by considering that additional work is done on the system by the feedback force. If more work is injected into the system, than heat can leave the system due to dissipation, the overall dissipative energy balance has to be negative.

\section{Concluding remarks}

In the present analysis we have comprehensively shown how the framework of  stochastic thermodynamics applies to Brownian motion at relativistic energies. Whereas cosmic gases and jets \cite{Meyer2018} may one day become the physical systems to be studied with this framework, there are experimental platforms that are somewhat ``closer to home''.

\subsection{Potential experimental significance}

In a rather recent experiment Pototsky \emph{et al.} \cite{Pototsky2012} demonstrated that relativistic Brownian motion can be studied in graphene. Graphene is essentially a single layer of graphite, i.e., carbon atoms in a honeycomb lattice. Its charge carriers have the remarkable property that they obey a linear dispersion relation, which means that the charge carriers behave like relativistic, massless Dirac fermions \cite{Wehling2014}. Pototsky \emph{et al.} \cite{Pototsky2012}  then demonstrated that for strong electric fields and at high temperatures the quantum signatures are suppressed, and that the dynamics of the charge carriers can be effectively described by the relativistic Langevin equation \eqref{eq:rel_Lang}. Our present results now open an avenue for the experimental study of relativistic stochastic thermodynamics in an already existing system.

\subsection{Summary}

In summary, we demonstrated that the hallmark results from stochastic thermodynamics, namely the fluctuation theorems with and without feedback hold at relativistic energies. To this end, we generalized stochastic energetics to the relativistic Langevin equation. We concluded that the expression for the stochastic work remains the same, whereas the stochastic heat accounts for the relativistic corrections for energy and momentum transfer between the traveling particle and a heat reservoir at rest. With this identification obtaining the fluctuation theorems without feedback turned out to be a straight forward exercise. On the other hand, relativistic systems experiencing feedback are more interesting, as due to the relativistic time dilation measurement and feedback action are not necessarily simultaneous for observer and relativistic particle. Thus, we argued that the framework of absolute irreversibility needs to be invoked to avoid formulating statements of the second law from acausal, i.e., unphysical events.

\invisiblesection{Acknowledgments}
\section*{Acknowledgments}
Elucidating discussions with J\"orn Dunkel on the nature of relativistic Brownian motion are gratefully acknowledged. This research was supported by grant number FQXi-RFP-1808 from the Foundational Questions Institute and Fetzer Franklin Fund, a donor advised fund of Silicon Valley Community Foundation.

\section*{References}


\providecommand{\newblock}{}

\end{document}